\documentclass[12pt]{article}
\usepackage{amsfonts,amssymb, amsmath,mathrsfs}

\def\on#1#2{\mathop{\vbox{\ialign{##\crcr\noalign{\kern2pt}
$\scriptstyle{#2}$\crcr\noalign{\kern2pt\nointerlineskip}
\kern-2pt$\hfil\displaystyle{#1}\hfil$\crcr}}}\limits}

\newtheorem{prop}{Proposition}

\newtheorem{remark}{Remark}
\newenvironment{rem}{\begin{remark} \rm}{\end{remark}}

\def\nn{ \nonumber }
\def\bq{ \begin{equation} }
\def\eq{ \end{equation} }
\def\ben{ \begin{eqnarray} }
\def\en{ \end{eqnarray} }
\def\frac#1#2{{#1\over #2}}
\def\dfrac#1#2{{\displaystyle{#1\over#2}}}

\def\e{{\rm e}}

\begin{document}

\title{A family of the Poisson brackets compatible with the Sklyanin bracket }
\author{
A. V. Tsiganov\\
St.Petersburg State University, St.Petersburg, Russia\\
\it\small e--mail: tsiganov@mph.phys.spbu.ru}

 \date{}
\maketitle

{\small
We introduce a family of compatible Poisson brackets
on the space of $2\times 2$ polynomial matrices, which contains the Sklyanin bracket,
and use it to derive a multi-Hamiltonian structure for a set of
integrable systems that includes $XXX$ Heisenberg magnet, the open and periodic Toda lattices,
the discrete self-trapping model and the Goryachev-Chaplygin gyrostat.}

\section{Introduction.}
\setcounter{equation}{0}
The ingenious discovery of Magri \cite{mag78,mag97} that integrable Hamiltonian
systems usually prove to be bi-Hamiltonian, and vice versa, leads us to the following fundamental problem: given a dynamical system which is Hamiltonian with respect to a  Poisson bracket $\{.,.\}_0$, how to find another Poisson bracket $\{.,.\}_1$ compatible with initial bracket and such that our system is Hamiltonian with respect to both brackets. This, along with the related problem of classification of compatible Poisson structures,
is nowadays a subject of intense research, see e.g. \cite{mag78,mag97,fp02,ts06c} and references therein.

In this paper we study a class of finite-dimensional Liouville integrable systems described
by the representations of the quadratic $r$-matrix Poisson algebra, or the Sklyanin algebra:
\bq
\{\,\on{T}{1}(\lambda),\,\on{T}{2}(\mu)\}= [r(\lambda-\mu),\,
\on{T}{1}(\lambda)\on{T}{2}(\mu)\,]\,, \label{rrpoi}
\eq
Here $\on{T}{1}(\lambda)=T(\lambda)\otimes \mathrm I\,,~\on{T}{2}(\mu)=\mathrm I\otimes T(\mu)$ and
$r(\lambda-\mu)$ is a classical $r$-matrix \cite{skl84}-\cite{skl92}.

The main result of the present paper is a family of the Poisson brackets $\{.,.\}_k$, which is compatible with the Sklyanin bracket (\ref{rrpoi}),  in the simplest case
of the $4\times4$ rational $r$-matrix
\bq
r(\lambda-\mu)=\dfrac{\eta}{\lambda-\mu}\Pi,\qquad \Pi=\left(\begin{array}{cccc}
 1 & 0 & 0 & 0 \\
 0 & 0 & 1 & 0 \\
 0 & 1 & 0 & 0 \\
 0 & 0 & 0 & 1
\end{array}\right)\,,\quad \eta\in {\mathbb C}\,,\label{rr}
\eq
and $2\times2$ matrix $T(\lambda)$, which depends polynomially on
the parameter $\lambda$
\ben\label{22T}
 T(\lambda)&=&\left(\begin{array}{cc}
 A (\lambda)& B (\lambda)\\
 C(\lambda) & D(\lambda)
\end{array}\right)\\
&=&\left(\begin{array}{ll}
\alpha\lambda^n+A_1\lambda^{n-1}+\ldots +A_n\qquad& \beta\lambda^n+B_1\lambda^{n-1}+\ldots+B_n \\
\gamma\lambda^n+ C_1\lambda^{n-1}+\ldots+C_n& \delta\lambda^n+D_1\lambda^{n-1}+\ldots+D_n
\end{array}\right).\nn
\en
The leading coefficients $\alpha,\beta,\gamma,\delta$ and $2n$ coefficients of the $\det T(\lambda)$
\bq
d(\lambda)=\mathrm{det}\,T(\lambda)=(\alpha\delta-\beta\gamma)\lambda^{2n}+Q_1\lambda^{2n-1}+\cdots+Q_{2n}\,.
\label{Acentre}
\eq
are Casimirs of the bracket (\ref{rrpoi}). Therefore, we have a $4n$-dimensional space of the coefficients $A_i, B_i, C_i$ and $D_i$ with $2n$ Casimir operators $Q_i$, leaving us with $n$ degrees of freedom.

For so-called open lattices independent Poisson involutive integrals of motion $H^o_i=A_i$, $i = 1,\ldots,n,$ are given by the coefficients of the entry $A(\lambda)$:
\bq\label{int-open}
A(\lambda) = \alpha\lambda^n +H^o_1\lambda^{n-1}+\cdots H^o_n,\qquad \{H^o_i,H^o_j\}=0\,.
\eq
In generic case integrals of motion are given by the coefficients of the tr$T(\lambda)$:
\bq\label{int-per}
\mbox{\rm tr} T(\lambda) = (\alpha+\delta)\lambda^n +H_1\lambda^{n-1}+\cdots H_n,\qquad
 \{H_i,H_j\}=0\,.
 \eq
These integrals of motion define two Liouville integrable systems, which are our generic models for the whole paper.
Bi-hamiltonian description of these models gives rise to the bi-hamiltonian description
of the Goryachev-Chaplygin gyrostat \cite{skl84}, open and periodic Toda lattice \cite{skl85}, inhomogeneous Heisenberg magnet \cite{skl92} and the discrete self-trapping (DST) model \cite{skl00}.

\section{The compatible bracket}
\setcounter{equation}{0}

In this section, we  describe the Poisson bracket compatible with the Sklyanin bracket.
The Poisson brackets $\{.,.\}_{0}$ and $\{.,.\}_{1}$ are compatible if every linear combination of them is still a Poisson bracket. The corresponding compatible Poisson tensors $P_0$ and $P_1$ satisfy to the following equations
\bq\label{sch-eq}
[\![P_0,P_0]\!]=[\![P_0,P_1]\!]=[\![P_1,P_1]\!]=0,
\eq
where $[\![.,.]\!]$ is the Schouten bracket \cite{fp02,mag78,mag97}. Remind that
on a smooth finite-dimensional manifold $\mathscr M$
the Schouten bracket of two bivectors  $X$ and $Y$  is an antisymmetric contravariant tensor of rank three and its components in local coordinates  $z_m$ read
\[
[\![X,Y]\!]^{ijk}=-\sum_{m=1}^{dim\, \mathscr M} \left(
X^{mk}\dfrac{\partial Y^{ij}}{\partial z_m}+Y^{mk}\dfrac{\partial
X^{ij}}{\partial z_m}+\mbox{cycle}(i,j,k)\, \right).
\]

\subsection{Open lattices}
The Sklyanin bracket (\ref{rrpoi}) amounts to having the following Poisson brackets between the entries $A(\lambda)$, $B(\lambda)$, $C(\lambda)$ and $D(\lambda)$ of the matrix $T(\lambda)$:
\bq\label{3}
\begin{array}{l}
\left\{A(\lambda),A(\mu)\right\}_0=\{B(\lambda),B(\mu)\}_0=\{C(\lambda),C(\mu)\}_0=\{D(\lambda),D(\mu)\}_0=0,
\\ \\
\left\{B(\lambda),A(\mu)\right\}_0=\dfrac{\eta}{\lambda-\mu}\Bigl(B(\lambda)A(\mu)-B(\mu)A(\lambda)\Bigr),
\\ \\
\left\{C(\lambda),A(\mu)\right\}_0=\dfrac{-\eta}{\lambda-\mu}\Bigl(C(\lambda)A(\mu)-C(\mu)A(\lambda)\Bigr),
\\ \\
\left\{B(\lambda),C(\mu)\right\}_0=\dfrac{\eta}{\lambda-\mu}\Bigl(D(\lambda)A(\mu)-D(\mu)A(\lambda)\Bigr).
\\ \\
\left\{B(\lambda),D(\mu)\right\}_0=\dfrac{-\eta}{\lambda-\mu}\Bigl(B(\lambda)D(\mu)-B(\mu)D(\lambda)\Bigr),
\\ \\
\left\{C(\lambda),D(\mu)\right\}_0=\dfrac{\eta}{\lambda-\mu}\Bigl(C(\lambda)D(\mu)-C(\mu)D(\lambda)\Bigr),
\\ \\
\left\{A(\lambda),D(\mu)\right\}_0=\dfrac{\eta}{\lambda-\mu}\Bigl(C(\lambda)B(\mu)-C(\mu)B(\lambda)\Bigr),
\end{array}
\eq
In (\ref{rrpoi}) matrix $r(\lambda-\mu)$ satisfies  the Yang-Baxter equation,
which ensures the Jacobi identity for the  brackets (\ref{3}).

\begin{prop}
The Sklyanin bracket (\ref{rrpoi}),(\ref{3}) is compatible with the following bracket $\{.,.\}_1$:
\bq\label{31}
\begin{array}{l}
\left\{A(\lambda),A(\mu)\right\}_1=\{B(\lambda),B(\mu)\}_1=\{C(\lambda),C(\mu)\}_1=0,
\\ \\
\left\{B(\lambda),A(\mu)\right\}_1=\dfrac{\eta}{\lambda-\mu}\Bigl(\lambda B(\lambda)A(\mu)-\mu B(\mu)A(\lambda)\Bigr)-\dfrac{\eta\,\beta}{\alpha}A(\lambda)A(\mu),
\\ \\
\left\{C(\lambda),A(\mu)\right\}_1=\dfrac{-\eta}{\lambda-\mu}\Bigl(\lambda C(\lambda)A(\mu)-\mu C(\mu)A(\lambda)\Bigr)+\dfrac{\eta\,\gamma}{\alpha}A(\lambda)A(\mu),
\\ \\
\left\{B(\lambda),C(\mu)\right\}_1=\dfrac{\eta}{\lambda-\mu}\Bigl(\lambda D(\lambda)A(\mu)-\mu D(\mu)A(\lambda)\Bigr)-\dfrac{\eta\,\delta}{\alpha}A(\lambda)A(\mu),
\\ \\
\left\{B(\lambda),D(\mu)\right\}_1=\dfrac{-\eta\,\lambda}{\lambda-\mu}\Bigl(B(\lambda)D(\mu)-B(\mu)D(\lambda)\Bigr)+
\eta A(\lambda)\left(\dfrac{\beta}{\alpha}D(\mu)-\dfrac{\delta}{\alpha}B(\mu)\right),
\\ \\
\left\{C(\lambda),D(\mu)\right\}_1=\dfrac{\eta\,\lambda}{\lambda-\mu}\Bigl(C(\lambda)D(\mu)-C(\mu)D(\lambda)\Bigr)
-\eta A(\lambda)\left(\dfrac{\gamma}{\alpha}D(\mu)-\dfrac{\delta}{\alpha}C(\mu)\right),
\\ \\
\left\{A(\lambda),D(\mu)\right\}_1=\dfrac{\eta\,\lambda}{\lambda-\mu}\Bigl(C(\lambda)B(\mu)-C(\mu)B(\lambda)\Bigr)
-\eta A(\lambda)\left(\dfrac{\gamma}{\alpha}B(\mu)-\dfrac{\beta}{\alpha}C(\mu)\right),
\end{array}
\eq
and
\ben
\left\{D(\lambda),D(\mu)\right\}_1&=&
\frac{\eta\,\gamma}{\alpha}\Bigl(D(\lambda)B(\mu)-D(\mu)B(\lambda)\Bigr)
-\frac{\eta\,\beta}{\alpha}\Bigl(D(\lambda)C(\mu)-D(\mu)C(\lambda)\Bigr)\nn\\
&+&\frac{\eta\,\delta}{\alpha}\Bigl(B(\lambda)C(\mu)-B(\mu)C(\lambda)\Bigr)\,.
\label{41}
\en
\end{prop}
\textbf{Proof}:
It is sufficient to check the statement on an open dense subset of the Sklyanin algebra defined by the assumption that $A(\lambda)$ and $B(\lambda)$ are co-prime and all roots of $A(\lambda)$ are distinct.

This assumption allows us to construct a separation representation for the Sklyanin algebra (\ref{rrpoi}). In this special representation one has $n$ pairs of Darboux variables, $\lambda_i$, $\mu_i$, $i=1,\ldots,n$, having the standard Poisson brackets,
\bq
\left\{\lambda_i,\lambda_j\right\}_0=\left\{\mu_i,\mu_j\right\}_0=0,\qquad \left\{\lambda_i,\mu_j\right\}_0=\delta_{ij},
\label{Darb}
\eq
with the $\lambda$-variables being $n$ zeros of the polynomial $A(\lambda)$ and the $\mu$-variables being values of the polynomial $B(\lambda)$ at those zeros,
\bq
A(\lambda_i)=0,\qquad \mu_i=\eta^{-1}\ln B(\lambda_i),\qquad i=1,\ldots,n.
\label{dn-var}
\eq
The interpolation data (\ref{dn-var}) plus $n$ identities
\[B(\lambda_i)C(\lambda_i)=-d(\lambda_i)\]
allow us to construct the needed separation representation for the whole algebra:
\bq
\label{DN-rep}
\begin{array}{l}
A(\lambda)=\alpha(\lambda-\lambda_1)(\lambda-\lambda_2)\cdots(\lambda-\lambda_n),\\
\\
B(\lambda)=A(\lambda)\left(\dfrac{\beta}{\alpha}+\sum_{i=1}^n \dfrac{\e^{\eta \mu_i}}{(\lambda-\lambda_i)A^{'}(\lambda_i)}\right),\\
\\
C(\lambda)=A(\lambda)\left(\dfrac{\gamma}{\alpha}-\sum_{i=1}^n \dfrac{d(\lambda_i)\;\e^{-\eta \mu_i}}
{(\lambda-\lambda_i)A^{'}(\lambda_i)}\right),\\
\\
D(\lambda)=\dfrac{d(\lambda)+B(\lambda)C(\lambda)}{A(\lambda)}\,.
\end{array}
\eq
The coefficients of the determinant
$d(\lambda)$ (\ref{Acentre})
are Casimir elements for the both brackets $\{.,.\}_0$ and $\{.,.\}_1$
and, therefore, we can easy calculate the bracket $\{.,.\}_1$ (\ref{31})--(\ref{41}) in ($\lambda,\mu$)-variables
\bq
\left\{\lambda_i,\lambda_j\right\}_1=\left\{\mu_i,\mu_j\right\}_1=0,\qquad \left\{\lambda_i,\mu_j\right\}_1=\lambda_i\delta_{ij},
\label{NDarb}
\eq
In order to complete the proof we have to check that brackets (\ref{NDarb}) is compatible with the canonical brackets (\ref{Darb}).
The compatibility of the brackets (\ref{Darb}),(\ref{NDarb}) implies the compatibility of the brackets (\ref{3}),(\ref{31}) and vice versa.

 \vskip0.3truecm

The ($\lambda,\mu$)-variables (\ref{dn-var}) are so-called special
Darboux-Nijenhuis coordinates \cite{mag78,mag97,fp02} because
\[
P_0=\left(
 \begin{array}{cc}
 0 & \mathrm I \\
 -\mathrm I & 0
 \end{array}
 \right)\,,\qquad P_1=\left(
 \begin{array}{cc}
 0 & \mbox{\rm diag}(\lambda_1,\ldots,\lambda_n) \\
 -\mbox{\rm diag}(\lambda_1,\ldots,\lambda_n) & 0
 \end{array}
 \right)\,,\qquad
\]
and the corresponding recursion operator $N$ takes the diagonal form
\bq\label{xy-pedr}
 N=P_1P_0^{-1}=\sum_{i=1}^n\lambda_i\left(
\dfrac{\partial }{\partial \lambda_i}\otimes d\lambda_i+ \dfrac{\partial
}{\partial \mu_i}\otimes d\mu_i\right).
\eq
These Poisson tensors $P_0$ and $P_1$ satisfy to the equations (\ref{sch-eq}) and the Nijenhuis torsion of $N$ vanishes as a consequence of the compatibility between $P_0$ and $P_1$.
\begin{prop}
Brackets (\ref{Darb}) and (\ref{NDarb}) between ($\lambda,\mu$)-variables belong to a
whole family of compatible Poisson brackets $\left\{.,.\right\}_k$ associated with the Poisson tensors
\[
P_k=N^kP_0=\left(
 \begin{array}{cc}
 0 & \mbox{\rm diag}(\lambda_1^k,\ldots,\lambda_n^k) \\
 -\mbox{\rm diag}(\lambda_1^k,\ldots,\lambda_n^k) & 0
 \end{array}
 \right)\,,\qquad k=0,\ldots,n.
\]
In the matrix form, these brackets are equal to
\ben
\left\{
\on{T(\lambda)}{1},\on{T(\mu)}{2}\right\}_{k}&=&r^{[k]}_{12}(\lambda,\mu)
\on{T(\lambda)}{1}\on{T(\mu)}{2}-
\on{T(\lambda)}{1}\on{T(\mu)}{2}r^{[k]}_{21}(\lambda,\mu)\nn\\
\nn\\
&+&\on{T(\lambda)}{1}s^{[k]}_{12}(\lambda,\mu)\on{T(\mu)}{2}-
\on{T(\mu)}{2}s^{[k]}_{21}(\lambda,\mu)\on{T(\lambda)}{1}\,. \label{br-skl2}
\en
Here
\bq
\label{rs-mat}
\begin{array}{ll}
r^{[k]}_{12}(\lambda,\mu)=\frac{\eta}{\lambda-\mu}\left(
 \begin{smallmatrix}
 1 & 0 & 0 & 0 \\
 0 & 1-\frac{\lambda^k+\mu^k}{2} & \mu^k & 0 \\
 0 & \lambda^k & 1-\frac{\lambda^k+\mu^k}{2} & 0 \\
 0 & \rho_C^{[k]} & -\rho_C^{[k]} & 1
 \end{smallmatrix}
 \right),\quad &
r^{[k]}_{21}(\lambda,\mu)=\frac{\eta}{\lambda-\mu}\left(
 \begin{smallmatrix}
 1 & 0 & 0 & 0 \\
 0 & 1-\frac{\lambda^k+\mu^k}{2} & \lambda^k & \rho_B^{[k]} \\
 0 & \mu^k & 1-\frac{\lambda^k+\mu^k}{2} & -\rho_B^{[k]} \\
 0 & 0 & 0 & 1
 \end{smallmatrix}
 \right) ,\\
\\
s^{[k]}_{12}(\lambda,\mu)=\frac{\eta}{\lambda-\mu}\left(
 \begin{smallmatrix}
 0 & \rho_B^{[k]} & 0 & 0 \\
 0 & \frac{\lambda^k-\mu^k}{2} & 0 & 0 \\
 \rho_C^{[k]} & \rho_D^{[k]} & \frac{\lambda^k-\mu^k}{2} & 0 \\
 0 & 0 & 0 & 0
 \end{smallmatrix}
 \right),\quad & s^{[k]}_{21}(\lambda,\mu)=\Pi s^{[k]}_{12}(\lambda,\mu)\Pi.
 \end{array}
\eq
and
\[
\rho_X^{[k]}=\dfrac{\lambda^kX(\lambda)}{A(\lambda)}-\dfrac{\mu^kX(\mu)}{A(\mu)}\,,
\qquad\mbox{\it where}\qquad X=B,C,D,
\]
is a difference of two polynomials, which are quotients of polynomials in  variables $\lambda$ and $\mu$ over a field.
\end{prop}
\textbf{Proof:} At $k=0$ one has $\rho_B^{[0]}=0$, $\rho_C^{[0]}=0$ and $\rho_D^{[0]}=0$, so
the bracket (\ref{br-skl2}) coincides with the Sklyanin bracket (\ref{rrpoi}). At $k=1$ we have
\[\rho_B^{[1]}=\frac{\beta(\lambda-\mu)}{\alpha},\qquad
\rho_C^{[1]}=\frac{\gamma(\lambda-\mu)}{\alpha},\qquad
\rho_D^{[1]}=\frac{\delta(\lambda-\mu)}{\alpha}\]
and bracket (\ref{br-skl2}) coincides with the bracket (\ref{31}).

At $k>1$
one can easily check that $k$-th brackets (\ref{br-skl2}) between polynomials $A(\lambda)$, $B(\lambda)$, $C(\lambda)$ and $D(\lambda)$ (\ref{DN-rep}) imply the brackets
\bq
\left\{\lambda_i,\lambda_j\right\}_k=\left\{\mu_i,\mu_j\right\}_k=0,\qquad \left\{\lambda_i,\mu_j\right\}_k=\lambda_i^k\delta_{ij}.
\label{NKDarb}
\eq
and vice versa. This completes the proof.
 \vskip0.3truecm

To proceed further we need to recall  that the normalized traces of the powers of $N$
\bq\label{mag-int}
{J}_m=\frac{1}{2m}\,\mbox{\rm trace}\,N^m=\sum_{i=1}^n\lambda_i^m,\qquad m=1,\ldots,n.
\eq
are integrals of motion satisfying Lenard-Magri
recurrent relations \cite{mag78,mag97}
\bq\label{len-mag} P_0dJ_1=0,\qquad
X_{{J}_i}=P_0d{J}_i=P_1d{J}_{i-1},\qquad P_1d{J}_n=0.
\eq
By definition (\ref{DN-rep}) polynomial
\[A(\lambda)=\alpha\lambda^n+A_1\lambda^{n-1}+\ldots +A_n=\alpha\prod_{i=1}^n(\lambda-\lambda_i)\]
is directly proportional to the minimal characteristic polynomial of  $N$ (\ref{xy-pedr})
\[\Delta_N(\lambda)=\left(\det(N-\lambda\mathrm I)\right)^{1/2}=\prod_{i=1}^n(\lambda-\lambda_i).
\]
Since Hamiltonians $H^o_i$ (\ref{int-open}) are related with integrals of
motion $J_m$ (\ref{mag-int}) by the triangular Newton formulas
\[
\alpha J_1=H^o_1,\quad \alpha J_2=H^o_2+\frac{(H^o_1
)^2}{2},\quad
\alpha J_3=H^o_3+H^o_2H^o_1+\frac{(H^o_1)^3}{3},\ldots.
\]
As a consequence of the recursion relations (\ref{len-mag}), the Hamiltonians
$H^o_i$, $i=1,\ldots,n$, satisfy the Fr\"obenius recursion relations
\bq\label{fr-ch1}
N^*\,dH^o_i=dH^o_{i+1}-\alpha^{-1}A_i\,dH^o_1,,
\eq
where $N^*=P_0^{-1}P_1$ and $H^o_{n+1}=0$.
Such as $A_i=H^o_i$ a straightforward computation shows that they are equivalent
\[
N^*\,dA(\lambda)=\lambda\,dA(\lambda)+A(\lambda)dA_1\,.
\]

The special Darboux-Nijenhuis coordinates $\lambda_i,\mu_i$ are variables of separation of the action-angle type \cite{fp02}, i.e. the corresponding separated equations are trivial
\[\{H^o_i,\lambda_j\}=\{J_i,\lambda_j\}=0\,,\qquad i,j=1,\ldots,n.\]
We can introduce another separated coordinates $u_i,v_i$, which are the so-called Sklyanin variables defined by
\[
B(u_i)=0,\qquad v_i=-\eta^{-1}\ln A(u_i)\,,\qquad i=1,\ldots,n.
\]
The separation representation of the algebra in $(u,v)$-variables has the form
\ben
B(\lambda)&=&\beta(\lambda-u_1)(\lambda-u_2)\cdots(\lambda-u_n),\nn\\
A(\lambda)&=&B(\lambda)\left(\frac{\alpha}{\beta}+\sum_{i=1}^n \frac{\e^{-\eta v_i}}{(\lambda-u_i)B'(u_i)}\right),\nn\\
D(\lambda)&=&B(\lambda)\left(\frac{\delta}{\beta}+\sum_{i=1}^n \frac{d(u_i)\;\e^{\eta v_i}}
{(\lambda-u_i)B'(u_i)}\right),\nn\\
C(\lambda)&=&\frac{A(\lambda)D(\lambda)-d(\lambda)}{B(\lambda)}\,.\nn
\en
Substituting matrix $T(\lambda)$ (\ref{22T}) with these entries
into the brackets $\{.,.\}_k$ (\ref{br-skl2}) at $k=0,1$ one gets that
$u_i,v_j$ coordinates are Darboux variables with respect to the Sklyanin bracket
\bq
\left\{u_i,u_j\right\}_0=\{v_i,v_j\}_0=0,\qquad \{u_i,v_j\}_0=\delta_{ij},
\label{Darb-skl}
\eq
whereas the second brackets look like
\[
\{u_i,u_j\}_1=0,\qquad
\{u_i,v_j\}_1=u_i\delta_{ij}-\frac{\beta\,A(u_j) }{\alpha\,B'(u_j)}\,,\qquad
\{v_i,v_j\}=\frac{A'(u_i)}{B'(u_i)}-\frac{A'(u_j)}{B'(u_j)}.
\]
The corresponding separated equations
\bq\label{toda-eqm}
\{A(\lambda),u_j\}_{k}=\lambda^k\,A(u_j)\,\prod_{i\neq
j}^{n-1}\dfrac{\lambda-u_i}{u_j-u_i}\,,\qquad j=1,\ldots,n.
\eq
are
linearized by the Abel transformation on the algebraic curve defined by $\e^{-\eta v_i}=A(u_i)$, see \cite{skl85,fl76,smirn98} and references within.

The special Darboux-Nijenhuis coordinates are dual to the Sklyanin variables.
Namely, $\lambda_i,\mu_i$ are roots of polynomial $A(\lambda)$ and values of polynomial $B(\lambda)$ at $\lambda=\lambda_i$, while $u_i,v_i$ are roots of polynomial $B(\lambda)$ and values of polynomial $A(\lambda)$ at
$\lambda=u_i$.

\subsection{Generic model}
There are many other Poisson brackets compatible with the standard one (\ref{Darb}).
The main property of the proposed above bracket $\{.,.\}_1$ (\ref{31})--(\ref{41}) is that
\[\left\{A(\lambda),A(\mu)\right\}_0=\{A(\lambda),A(\mu)\}_1=0\,.\]
It ensures that integrals of motion $H^o_i$ for the open lattices are in bi-involution
\[\left\{H^o_i,H_j^o\right\}_0=\left\{H^o_i,H^o_j\right\}_1=0.\]

In this subsection we are looking for bracket $\{.,.\}_1^{\prime}$, which has to guarantee the similar property for generic integrals of motion $H_i$ (\ref{int-per}) from tr$T(\lambda)$
\[\left\{H_i,H_j\right\}_0=\left\{H_i,H_j\right\}^{\prime}_1=0,\qquad i,j=1,\ldots,n.\]
Remind that  $\{.,.\}_0$ is the Sklyanin bracket (\ref{rrpoi}), which already has the necessary property
\[
\left\{\mbox{\rm tr} T(\lambda),\mbox{\rm tr} T(\mu)\right\}_0=0\,.
\]
The following propositions can be ascertained by means of direct calculations.
\begin{prop}\label{prop3}
If $\alpha=\delta$ and $\beta=\gamma=0$ in $T(\lambda)$ (\ref{22T}), then
\bq\label{bi-Gen1}
\left\{\mbox{\rm tr} T(\lambda),\mbox{\rm tr} T(\mu)\right\}_1=0\,.
\eq
\end{prop}
So, the desired bracket $\{.,.\}_1^{\prime}$ may be obtained from the bracket $\{.,.\}_1$ (\ref{31})--(\ref{41}) by using special canonical transformations, which are generated by the suitable transformations of the matrix $T(\lambda)$.

\begin{prop}
The Sklyanin bracket (\ref{rrpoi}) is invariant with respect to transformation
\bq\label{tr-T}
T(\lambda)\to V_1\,T(\lambda)\,V_2,\qquad V_i=\left(
 \begin{array}{cc}
 \alpha_i & \beta_i \\
 \gamma_i & \delta_i
 \end{array}
 \right),\qquad \alpha_i,\beta_i,\gamma_i,\delta_i\in\mathbb R\,,
\eq
where $V_{1,2}$ are numerical matrices. If \[\beta_1\gamma_2+\delta_1\delta_2=0,\]
the bracket (\ref{31})--(\ref{41}) after transformation (\ref{tr-T}) has the necessary property
\bq\label{bi-Gen}
\left\{\mbox{\rm tr} T(\lambda),\mbox{\rm tr} T(\mu)\right\}^{\prime}_1=0\,.
\eq
\end{prop}
We present an explicit form of the bracket $\{.,.\}_1^{\prime}$ in the Section 4 devoted to the periodic Toda lattice.

\section{The Heisenberg magnet}
\setcounter{equation}{0}
Another important representation of the quadratic algebra with the generators
$A_{i},B_{i},$ $C_{i}$ and $D_{i}$ comes as a consequence of the
co-multiplication property of the Sklyanin algebra (\ref{rrpoi}). Essentially, it means that the matrix $T(\lambda)$ (\ref{22T}) can be factorized into a product of elementary matrices, each containing only one degree of freedom. In this picture, our main model turns out to be an $n$-site Heisenberg magnet, which is an integrable lattice of $n$ sl(2) spins with nearest neighbor interaction.

In the lattice representation the matrix $T(\lambda)$ (\ref{22T}) acquires the following form:
\bq\label{s-mat}
T(\lambda)=L_1(\lambda-c_1)\;L_{2}(\lambda-c_{2})\;\cdots\;L_n(\lambda-c_n)\;
,
\eq
with
\bq
L_m(\lambda)=\left(\begin{matrix}
\lambda-s_3^{(m)}& s_1^{(m)}+\mathrm{i}s^{(m)}_2\\
s_1^{(m)}-\mathrm{i}s^{(m)}_2&\lambda+s_3^{(m)}
\end{matrix}\right),\qquad m=1,\ldots,n.
\eq
Here $s_3^{(m)}$ are dynamical variables, $c_m$ are arbitrary numbers and $\mathrm i=\sqrt{-1}$\,.

Substituting matrix (\ref{s-mat}) into the Sklyanin bracket (\ref{rrpoi})
 and brackets (\ref{31})-(\ref{41}) at $\eta=\mathrm{i}$
 one gets canonical brackets on the direct sum of sl(2)
\bq \label{P-sl2}
\left\{s_i^{(m)},s_j^{(m)}\right\}_0=\varepsilon_{ijk}\,s_k^{(m)}\,,
\eq
and second compatible brackets
\[
\left\{s_i^{(m)},s_j^{(m)}\right\}_1=\varepsilon_{ijk}\,s_k^{(m)}\bigl(c_m-s_3^{(m)}\bigr),\qquad
\left\{s_i^{(m)},s_j^{(\ell)}\right\}_1=\left(P_1^{(m\ell)}\right)_{ij}\,,\quad m\neq\ell,
\]
where $\varepsilon_{ijk}$ is the totally skew-symmetric tensor and
\[P_1^{(m\ell)}=\left(
\begin{array}{ccr}
-\mathrm{i}\bigl(s^{(m)}_3 s^{(\ell)}_3+s^{(m)}_2 s^{(\ell)}_2\bigr)&
\mathrm{i}s^{(m)}_2 s^{(\ell)}_1 -s^{(m)}_3 s^{(\ell)}_3&
\mathrm{i} s^{(m)}_3 \bigl(s^{(\ell)}_1-\mathrm{i} s^{(\ell)}_2\bigr) \\
\\
\mathrm{i}s^{(m)}_1 s^{(\ell)}_2 +s^{(m)}_3 s^{(\ell)}_3&
-\mathrm{i}\bigl(s^{(m)}_3 s^{(\ell)}_3+s^{(m)}_1 s^{(\ell)}_1\bigr)&
-s^{(m)}_3 \bigl(s^{(\ell)}_1-\mathrm{i}s^{(\ell)}_2\bigr)\\
\\
-\mathrm{i}\bigl(s^{(m)}_1+\mathrm{i}s^{(m)}_2\bigr)s^{(\ell)}_3&
\bigl(s^{(m)}_1+\mathrm{i}s^{(m)}_2\bigr)s^{(\ell)}_3 &
-\mathrm{i}\bigl(s^{(m)}_1+\mathrm{i}s^{(m)}_2\bigr)\bigl(s^{(\ell)}_1-\mathrm{i}s^{(\ell)}_2\bigr)
\end{array}\right).
\]
The corresponding Poisson tensors $P_0$ and $P_1$ are degenerate and, therefore,
the Hamiltonians $H^o_i$ satisfy
the Fr\"obenius recurrence relations (\ref{fr-ch1}) in the following form
\bq\label{hm-fr}
{P}_1dH_i^o={P}_0\left(dH_{i+1}^o-A_idH_1^o\right)\,,\qquad i=1,\ldots,n,
\eq
where $H^o_{n+1}=0$ and $A_i=H_i^o$ are coefficients of the polynomial
$A(\lambda)$. The first integrals of motion are
\[
H_1^o=\sum_{m=1}^n (c_m-s^{(m)}_3)\,,\qquad
H_2^o=\sum_{m>\ell} \bigl(s^{(m)}_1-\mathrm{i}s^{(m)}_2\bigr)\bigl(s^{(\ell)}_1+\mathrm{i}s^{(\ell)}_2\bigr)-\frac12\sum_{m=1}^n (c_m-s^{(m)}_3)^2
+\frac{\left(H_1^o\right)^2}2\,.
\]
Such as $\alpha=\delta$ and $\beta=\gamma=0$ we can use these brackets for the open and periodic lattices simultaneously. It means that Hamiltonians $H_i$ (\ref{int-per}) from the tr${T}(\lambda)$
satisfy the Fr\"obenius equations (\ref{hm-fr}) too.

\section{The Toda lattices}
\setcounter{equation}{0}

The Toda lattices appear as a specialization of our basic model when the parameters are fixed as follows:
\bq
\beta=\gamma=\delta=0\qquad \mbox{\rm and}\qquad \det T(\lambda)=1.
\eq
We also put $\alpha=1$ and $\eta=-1$.
In the lattice representation, the monodromy matrix $T$ (\ref{22T}) acquires
the form
\bq\label{22toda}
T(\lambda)=L_1(\lambda)\cdots L_{n-1}(\lambda)\,L_n(\lambda)
\,, \qquad
L_i=\left(\begin{array}{cc}
 \lambda-p_i &\, -\e^{q_i} \\
 \e^{-q_i}& 0
\end{array}\right)\,.
\eq
Here $p_i,q_i$ are dynamical variables.

\subsection{Open lattice}
Substituting matrix $T(\lambda)$ (\ref{22toda}) into the brackets $\{.,.\}_k$ (\ref{br-skl2}) at $k=0,1$
one gets that the Poisson tensors $P_0$ and $P_1$ in $(p,q)$ variables take the form
\ben\label{P-toda}
P_0&=&\sum_{i=1}^n \dfrac{\partial}{\partial q_{i}}\wedge\dfrac{\partial}{\partial
p_{i}}\,,\\
\nn\\
P_1&=&\sum_{i=1}^{n-1}
\e^{q_i-q_{i+1}}\dfrac{\partial}{\partial
p_{i+1}}\wedge\dfrac{\partial}{\partial p_{i}} +\sum_{i=1}^n
p_i\dfrac{\partial}{\partial q_{i}}\wedge\dfrac{\partial}{\partial
p_{i}}+\sum_{i<j}^n \dfrac{\partial}{\partial
q_{j}}\wedge\dfrac{\partial}{\partial q_{i}}.\nn
\en
Namely this bi-hamiltonian structure of the open Toda lattice
was obtained in \cite{das89}.

For the open Toda lattice the Hamiltonians $H^o_i$ from the
$A(\lambda)=\lambda^{n}+H^o_{1}\lambda^{n-1}+\ldots+H^o_n$
satisfy the Fr\"obenius relations (\ref{fr-ch1}). The first integrals of motion are equal to
\bq
H_1^o=-\sum_{i=1}^np_i,\qquad
H_2^o=\dfrac12\sum_{i=1}^n {p_i}^2+\sum_{i=1}^{n-1} \e^{q_i-q_{i+1}}-\dfrac{1}{2}\left(\sum_{i=1}^np_i\right)^2\,.
\eq

The Sklyanin variables $u_i, v_i$ are introduced as before:
\bq\label{uv-Toda1}
B(u_i)=0,\qquad v_i=-\eta^{-1}\ln A(u_i)\,,\qquad i=1,\ldots,n-1,\eq
the only difference now is that this gives only $n-1$
instead of $n$ separation pairs. The missing pair of
canonical variables is defined as follows:
\bq\label{uv-Toda2}
v_n=\ln\, b_1 =-q_n,\qquad u_n=-a_1=\sum_{i=1}^np_i.
\eq
The separation representation of the algebra in $(u,v)$-variables may be found in \cite{skl85,ts06a}.
It is easy to prove \cite{ts06a} that $(u,v)$-variables are Darboux variables
\[\omega=P_0^{-1}=\sum_{i=1}^n du_i\wedge dv_i,\]
and the only nonzero second Poisson brackets are
\[
\begin{array}{lll}
\left\{u_j,v_i\right\}_1=u_i\,\delta_{ij},\quad&
\left\{u_n,u_i\right\}_1= -\e^{-v_n}\frac{A(u_i)}{B'(u_i)},
\quad&
\left\{u_n,v_i\right\}_1=-\e^{-v_n}\frac{A'(u_i)}{B'(u_i)}\,,\\
\\
\left\{v_n,v_i\right\}_1=-1,\quad&\left\{u_n,v_n\right\}_1=- \displaystyle{\sum_{i=1}^{n}} u_i. \quad&
\end{array}
\]

\begin{rem}
From the factorization (\ref{22toda}) of the monodromy
matrix $T(\lambda)$ one gets
\[
 B_{n}(\lambda)=-\e^{q_{n}}A_{n-1}(\lambda)\,\qquad \Rightarrow\qquad B_{n}(u_j)=-\e^{q_{n}}A_{n-1}(\lambda_j)=0.
\]
This implies that for the $(n-1)$-particle chain special
Darboux-Nijenhuis variables $\lambda_j$ coincide with the Sklyanin variables $u_j$, $i=1,\ldots,n-1$ for the $n$-particle chain.
\end{rem}

\subsection{Periodic lattice}
For the Toda lattice $\alpha\neq\delta$ and, therefore, in order to get new bracket $\{.,.\}_1'$ with the the necessary property (\ref{bi-Gen}) we have to apply transformation
(\ref{tr-T}) to the initial bracket $\{.,.\}_1$ (\ref{31})-(\ref{41}). If we put
\[
V_1=\left(
 \begin{array}{cc}
 1 & -1 \\
 0 & 1
 \end{array}
 \right),\qquad\mbox{\rm and}\qquad V_2=\left(
 \begin{array}{cc}
 1 & 0 \\
 1 & 1
 \end{array}
 \right)\,,
\]
then one gets the following brackets between the entries of $T(\lambda)$:
\bq\label{32}
\begin{array}{ll}
\{A(\lambda),A(\mu)\}_1^{\prime}=\eta\bigl(B(\lambda)C(\mu)-B(\mu)C(\lambda)\bigr),\quad& \{D(\lambda),D(\mu)\}_1^{\prime}=0,\\
\\
\{A(\lambda),D(\mu)\}_1^{\prime}=\dfrac{\eta\lambda}{\lambda-\mu}\bigl(C(\lambda)B(\mu)-C(\mu)B(\lambda)\bigr)
\\ \\
\{B(\lambda),B(\mu)\}_1^{\prime}=\eta\bigl(B(\lambda)D(\mu)-B(\mu)D(\lambda)\bigr),\quad&
\\ \\
\{C(\lambda),C(\mu)\}_1^{\prime}=\eta\bigl(C(\lambda)D(\mu)-C(\mu)D(\lambda)\bigr),&\\
\\
\{D(\lambda),B(\mu)\}_1^{\prime}=\dfrac{\eta\mu}{\lambda-\mu}\bigl(B(\lambda)D(\mu)-B(\mu)D(\lambda)\bigr)\\
\\
\{D(\lambda),C(\mu)\}_1^{\prime}=\dfrac{-\eta\mu}{\lambda-\mu}
\bigl(C(\lambda)D(\mu)-C(\mu)D(\lambda)\bigr)
\end{array}
\eq
and
\bq\label{42}
\begin{array}{l}
\{A(\lambda),B(\mu)\}_1^{\prime}=
\frac{-\eta\bigl(\lambda A(\mu)B(\lambda)-\mu A(\lambda)B(\mu)\bigr)}{\lambda-\mu}
+\eta\Bigl(B(\lambda)D(\mu)+\bigl(B(\lambda)-C(\lambda)\bigr)B(\mu)\Bigr)\,,
\\
 \\
 \{A(\lambda),C(\mu)\}_1^{\prime}=\frac{\eta\bigl(\lambda A(\mu)C(\lambda)-\mu A(\lambda)C(\mu)}{\lambda-\mu}
-\eta\Bigl(C(\lambda)D(\mu)+\bigl(B(\lambda)-C(\lambda)\bigr)C(\mu)\Bigr),
 \\
 \\
 \{B(\lambda),C(\mu)\}_1^{\prime}=
\frac{\eta\bigl(\lambda A(\mu)D(\lambda)-\mu A(\lambda)D(\mu)\bigr)}{\lambda-\mu}
-\eta\bigl(B(\lambda)D(\mu)-D(\lambda)C(\mu)+D(\lambda)D(\mu)\bigr)\,.
\end{array}
 \eq
 According to proposition 4 these brackets have the necessary property (\ref{bi-Gen}).

Substituting matrix $T(\lambda)$ (\ref{22toda}) into the brackets (\ref{32})-(\ref{42})
one gets that the Poisson tensor $P'_1$ in $(p,q)$ variables takes the form
\ben\label{Pp-toda}
P'_1&=&\sum_{i=1}^{n-1}
\e^{q_i-q_{i+1}}\dfrac{\partial}{\partial
p_{i+1}}\wedge\dfrac{\partial}{\partial p_{i}} +\sum_{i=1}^n
p_i\dfrac{\partial}{\partial q_{i}}\wedge\dfrac{\partial}{\partial
p_{i}}+\sum_{i<j}^n \dfrac{\partial}{\partial
q_{j}}\wedge\dfrac{\partial}{\partial q_{i}}\nn\\
&-&\sum_{i=1}^n \left(\e^{q_1}\dfrac{\partial}{\partial
q_{i}}\wedge\dfrac{\partial}{\partial p_{1}}
+\e^{q_n}\dfrac{\partial}{\partial
q_{i}}\wedge\dfrac{\partial}{\partial p_{n}}
\right)\,.\nn
\en
For the periodic Toda lattice the Hamiltonians $H_1$ and $H_2$ from the
tr$T(\lambda)=\lambda^{n}+H_{1}\lambda^{n-1}+\ldots+H_0$ are equal to
\bq
H_1=H^o_1=-\sum_{i=1}^np_i,\qquad
H_2=H^o_2+\e^{q_n-q_1}=\dfrac12\sum_{i=1}^n {p_i}^2+\sum_{i=1}^{n} \e^{q_i-q_{i+1}}-\dfrac{1}{2}\left(\sum_{i=1}^np_i\right)^2\,,
\eq
where $q_{n+i}=q_i$. These Hamiltonians $H_i$, $i=1,\ldots,n$, form the Fr\"obenius chain
\bq\label{fr-ch2}
{N}^*dH_i=dH_{i+1}+c_idH_1,\qquad \mbox{\rm with}\qquad H_{n+1}=0\,.
\eq
Here $N^*=P_0^{-1}P'_1$ and $c_i$ are coefficients of the minimal characteristic polynomial of the recursion operator
\bq\label{c-period}
\Delta_N(\lambda)=\Bigl(\det(N-\lambda\, {\mathrm
I})\Bigr)^{1/2}=\lambda^n-(c_1\lambda^{n-1}+\cdots+c_n)\,,
\eq
which can be defined directly via the entries of the matrix $T(\lambda)$
\bq\label{det-period}
\Delta_N(\lambda)=A(\lambda)+B(\lambda)-C(\lambda)-D(\lambda)\,.\eq

\begin{rem}
Transformations (\ref{tr-T}) of the matrix $T(\lambda)$ give rise to canonical transformations in the phase space. As sequence tensor $P'_1$ (\ref{Pp-toda}) coincides with tensor $P_1$ (\ref{P-toda}) after the following canonical transformation
\[
p_1\to p_1+\e^{-q_1},\qquad p_n\to p_n+\e^{q_n}\,,
\]
which identifies coefficients $c_i$ with integrals of motion
for the open Toda lattice $c_i=-H_i^o$.
\end{rem}

\section{Integrable DST model}
\setcounter{equation}{0}

The integrable case of the DST (discrete self-trapping) model
with $n$ degrees of freedom was
studied in \cite{skl00}. It appears as a specialization of our
basic model when several parameters vanish:
\bq
\beta=\gamma=\delta=0\qquad \mbox{\rm and}\qquad Q_j=0, \qquad j=1,\ldots,n-1.
\eq
We also put $\alpha=1$ and $\eta=-1$. In the lattice representation, the matrix $T(\lambda)$ (\ref{22T}) acquires
the form
\bq\label{dst-mat}
T(\lambda)=L_1(\lambda-c_1)\;L_{2}(\lambda-c_{2})\;\cdots\;L_n(\lambda-c_n),
\quad\mbox{\rm with}\quad
L_i(\lambda)=\left(\begin{matrix}
\lambda-q_ip_i& bq_i\cr -p_i&b
\end{matrix}\right).
\eq
Here $p_i,q_i$ are dynamical variables, whereas $b$ and $c_i$ are numbers entering into
the Casimir function (\ref{Acentre})
\[
d(\lambda)=\det T(\lambda)=b^n(\lambda-c_1)(\lambda-c_2)\cdots (\lambda-c_n).
\]
Substituting matrix $T(\lambda)$ (\ref{dst-mat}) into the Sklyanin bracket (\ref{rrpoi}) and into
the brackets (\ref{32})-(\ref{42}) one gets canonical brackets
\[
\left\{p_i,q_j\right\}_0=\delta_{ij},\qquad \left\{q_i,q_j\right\}_0=\left\{p_i,p_j\right\}_0=0,\qquad i,j=1,\ldots,n\,.
\]
and quadratic brackets
\ben
\left\{q_i,q_j\right\}_1&=-Q_iQ_j,\qquad
\left\{q_i,p_j\right\}_1&=Q_iP_j-c_i\,\delta_{ij},\qquad i>j\nn\\
\left\{p_i,p_j\right\}_1&=-P_iP_j,
\qquad
\left\{p_i,q_j\right\}_1&=q_ip_j-b\,\delta_{i+1j}\,,\nn
\en
where $Q_1=q_1+1$, $P_n=p_n+b$ and $Q_i=q_i$, $P_i=p_i$ for other values of index $i$.

As above the Hamiltonians $H_i$, $i=1,\ldots,n$, from the
tr$T(\lambda)=\lambda^{n}+H_{1}\lambda^{n-1}+\ldots+H_n$ satisfy the Fr\"obenius relations (\ref{fr-ch2}). The two first Hamiltonians of the system are
\ben
H_1&=&-\sum_{i=1}^n(q_ip_i-c_i),\\
H_2&=&\sum_{i>j}(q_ip_i-c_i)(q_jp_j-c_j)
-b\sum_{i=1}^nq_{i}p_{i+1},\qquad p_{n+1}\equiv p_1.
\nonumber\en
The Sklyanin variables $(u_i,v_i)$, $i=1,\ldots,n$, are introduced by the same
formulae as for the Toda lattice, cf. (\ref{uv-Toda1}) and (\ref{uv-Toda2}).

\section{The Goryachev-Chaplygin gyrostat}
\setcounter{equation}{0}

Let us consider the matrix $T(\lambda)$ introduced in \cite{kuzts89}
\bq\label{neu-mat}
T(\lambda)=
\left(
 \begin{array}{cc}
 \lambda^2-2\lambda J_3-J_1^2-J_2^2 & (x_1+\mathrm{i}x_2)\lambda-x_3(J_1+\mathrm{i}J_2) \\
 (x_1-\mathrm{i}x_2)\lambda-x_3(J_1-\mathrm{i}J_2) & -x_3^2
 \end{array}
\right)\,.
\eq
Substituting matrix (\ref{neu-mat}) into the Sklyanin bracket (\ref{rrpoi})
 and brackets (\ref{31})-(\ref{41}) at $\eta=2\mathrm{i}$ one gets canonical Poisson tensor on
 the dual space of Euclidean algebra $e(3)$
\bq\label{can-e3}
P_0= \left(\begin{matrix}
0& 0& 0& 0& x_3& -x_2\\
*& 0& 0& -x_3& 0& x_1\\
*& *& 0& x_2& -x_1& 0\\
*& *& *& 0& J_3& -J_2\\
*& *& *& *& 0& J_1\\
*& *& *& *& *& 0
\end{matrix}\right)
\eq
and the following quadratic tensor
\bq\label{sec-e3}
P_1=\left(\begin{matrix}
0& -x_3^2& x_3 x_2& - x_2J_1& -x_2 J_2& x_3 J_2-2x_2J_3\\
*& 0& -x_3 x_1& x_1J_1& x_1J_2 & 2x_1 J_3 -x_3 J_1\\
*& *& 0& 0& 0& -x_1J_2+x_2J_1\\
*& *& *& 0& -J_1^2-J_2^2& -J_3 J_2\\
*& *& *& *& 0& J_1 J_3\\
*& *& *& *& *& 0
\end{matrix}\right).
\eq
These tensors satisfy equations (\ref{sch-eq}) at any values of the Casimir functions
\[
\mathcal C_1=x_1^2+x_2^2+x_3^2,\qquad \mathcal C_2=x_1J_1+x_2J_2+x_3J_3\,.
\]
However, in the proposed method coefficients of $\det T=-\mathcal C_1\lambda^2+x_3\,\mathcal C_2\,\lambda$ have to be the Casimir functions and, therefore, we have to put $\mathcal C_2=0$. As sequence, we have $\{A(\lambda),A(\mu)\}_1=0$ at $\mathcal C_2=0$ only.

\begin{rem}
Solving equations $P_0dH^o_2=(P_1+\alpha P_0)H^o_1$ at arbitrary values of $\mathcal C_{1,2}$ one gets
\[
H^o_1=J_3,\qquad H^o_2=J_1^2+J_2^2+2J_3^2+\alpha J_3\,.
\]
Here $H^o_2$ is a kinetic part of the Hamiltonian for the
Kowalevski gyrostat, which may be studied by using $2\times 2$ Lax matrix
$L(\lambda)=K_+T(\lambda)K_-T^{-1}(-\lambda)$ \cite{kuzts89}.
The tensor $P_1$ (\ref{sec-e3}) differs from the Poisson tensor for the Kowalevski
gyrostat, which appears from the linear $r$-matrix algebra \cite{ts06c}.
\end{rem}

The $2\times 2$ Lax matrix for the Goryachev-Chaplygin gyrostat looks like \cite{skl84}
\bq
\widetilde{T}(\lambda)=
\left(\begin{array}{cc}
\e^{\frac{q}{2}}&0\\
0&\e^{-\frac{q}{2}}
\end{array}\right)
 \left(\begin{array}{cc}
 \lambda+2J_3+p & a \\
 a & 0 \\
 \end{array}
 \right)\,T(\lambda)\,
\left(\begin{array}{cc}
\e^{-\frac{q}{2}}&0\\
0&\e^{\frac{q}{2}}
\end{array}\right)\,.
\eq
Here $p,q$ are additional dynamical variables, $a$ is an arbitrary number and $T(\lambda)$ is given by (\ref{neu-mat}).

Substituting this matrix into the Sklyanin bracket (\ref{rrpoi}) and into
the brackets (\ref{32})-(\ref{42}) one gets the compatible Poisson tensors on the extended phase space $e^*(3)\ltimes (p,q)$
\bq\label{gch-can}
\widetilde{P}_0\equiv\left(\begin{array}{cc}P_0&W_0\\
W_0^T&G_0
\end{array}\right)=\left(
\begin{array}{c|c}
\begin{array}{c} \quad P_0 \quad\end{array} & \begin{array}{c}0\\ \vdots\end{array}\\
\hrulefill& \hrulefill\\
\begin{array}{c}\quad *\quad \end{array}&\begin{array}{cc}0&2\mathrm{i}\\
 *&0\end{array}
\end{array}
\right)
\eq
and
\bq\label{gch-sec}
\widetilde{P}_1\equiv\left(\begin{array}{cc}P_1&W_1\\
W_1^T&G_1
\end{array}\right)=\left(
\begin{array}{l|l}
\begin{array}{c} \qquad P_1\end{array}\qquad&
\begin{array}{rr}
-2x_3J_2+2px_2+8x_2J_3&-2x_2\\
2x_3J_1-2px_1-8x_1J_3& 2x_1\\
2x_1J_2-2x_2J_1& 0\\
2J_2(p+3J_3)& -2J_2+\mathrm{i}x_3\e^q\\
-2ax_3-2pJ_1-6J_1J_2& 2J_1-x_3\e^q\\
2ax_2& -\mathrm{i}(x_1+\mathrm{i}x_2)\e^q\end{array}\\
\hrulefill&\hrulefill\\
\begin{array}{c}\qquad *\end{array}\qquad&\begin{array}{rr}
0& 2\mathrm{i}\bigl((x_1-\mathrm{i}x_2)\e^q-2J_3-p-a\e^q\bigr)\\
*& 0\end{array}
\end{array}\right),
\eq
which satisfy equations (\ref{sch-eq}) at $\mathcal C_2=0$ only.
Here $P_0$ and $P_1$ are given by (\ref{can-e3}) and (\ref{sec-e3}).

The Hamiltonians $H_i$ from the tr$\widetilde{T}(\lambda)=\lambda^3+H_1\lambda+\lambda^2H_2+H_3$ are
\ben
H_1&=&p,\nn\\ H_2&=&-\left(J_1^2+J_2^2+4J_3^2+2pJ_3-2ax_1\right)\,,
\nn\\
H_3&=&-(2J_3+p)(J_1^2+J_2^2)-2ax_3J_1\,.\nn
\en
The obtained tensors $\widetilde{P}_0$ and $\widetilde{P}_1$ are degenerate and, therefore, the Hamiltonians $H_i$ reproduce the Fr\"obenius chain in the following form
\bq\label{gch-fr}
\widetilde{P}_1dH_i=\widetilde{P}_0\left(dH_{i+1}+c_idH_1\right)\,,\qquad i=1,2,3,
\eq
where $H_4=0$ and $c_i$ are coefficients  of the polynomial
$\Delta_N(\lambda)=A(\lambda)+B(\lambda)-C(\lambda)-D(\lambda)$ (\ref{c-period})-(\ref{det-period}).

At $p=\rho$ and $q=0$ matrices $G_0$ (\ref{gch-can}) and $G_1$ (\ref{gch-sec})
are (generically) non-degenerate. So, the Dirac procedure can reduce
pencil $\widetilde{P}_0+\lambda \widetilde{P}_1$ to a new Poisson pencil
$\widetilde{P}^D_0+\lambda \widetilde{P}^D_1$ on $e^*(3)$ defined by
\[\widetilde{P}_k^D=P_k+\left(W_k\,G_k^{-1}\,W_k^T\right)_{p=\rho,\,q=0}\,,\qquad k=0,1.\]
Here $P_0=\widetilde{P}_0^D$ is canonical Poisson tensor (\ref{can-e3}) and $P_1$ is given by (\ref{sec-e3}). This reduction procedure preserves equations (\ref{gch-fr})
for the reduced integrals of motion.

\section{Conclusion}

We present a family of compatible Poisson brackets (\ref{br-skl2}), that includes the Sklyanin bracket, and prove that the Sklyanin variables are dual to the special Darboux-Nijenhuis coordinates associated with these brackets.
The application of the $r$-matrix formalism is extremely useful here resulting in drastic reduction of the calculations for a whole set of integrable systems.

The construction can be generalized to other $r$-matrix algebras.
Remind, if one substitutes $T(\lambda)=1+\varepsilon L(\lambda)+O(\varepsilon^2)$, $r=\varepsilon\, r$ into (\ref{rrpoi}) and let $\varepsilon\to 0$ one gets a linear bracket. Then if $T(\lambda)$ satisfy
the Sklyanin bracket (\ref{rrpoi}), then the matrix $\mathcal T(\lambda)=T(\lambda)K_-T^{-1}(-\lambda)$ obeys to the reflection equation algebra \cite{skl88}. The corresponding compatible brackets for the open generalized Toda lattices was considered in \cite{ts06b}.

Moreover, the whole construction can immediately be transferred to the quantum case because $r$-matrices in (\ref{br-skl2}) became dynamical matrices at $k>1$ only.

We would like to thank I.V. Komarov and V.I. Inozemtsev for useful and interesting discussions. The research was partially supported by the RFBR grant 06-01-00140.


\begin{thebibliography}{10}
\bibitem{das89}
A. Das, S. Okubo,
\newblock{\em A systematic study of the Toda lattice},
Ann. Phys., v.30, p.215-232, 1989.


 \bibitem{fp02}
G. Falqui, M. Pedroni,
\newblock{\em Separation of variables for bi-Hamiltonian systems},
\newblock{Math. Phys. Anal. Geom.}, v.6, p.139-179, 2003.

\bibitem{fl76}
H. Flaschka, D. W. McLaughlin,
 \newblock{\em Canonically
conjugate variables for the Korteweg-de Vries equation and the Toda
lattice with periodic boundary conditions}, Progr. Theor. Phys.,
v.55, p.438-456, 1976.

\bibitem{kuzts89}
V.B. Kuznetsov and A.V. Tsiganov,
\newblock{\em A special case of Neumann's system and
the Kowalewski-Chaplygin-Goryachev top},
\newblock {J.Phys. A}, v.22, p.L73, 1989.

\bibitem{skl00}
V.B. Kuznetsov, M. Salerno and E.K. Sklyanin,
\newblock{\em Quantum B\"acklund transformation for the integrable DST model.},
 J. Phys. A, v.33, p. 171-189, 2000.

\bibitem{mag78}
F. Magri,
\newblock{\em A simple model of the integrable Hamiltonian equation},
 J. Math. Phys., v.19, p.1156-1162, 1978.

 \bibitem{mag97}
F. Magri,
\newblock{\em Eight lectures on Integrable Systems},
 {Lecture Notes in Physics, Springer Verlag, Berlin-Heidelberg}, v.495, p.256-296, 1997.

\bibitem{skl84}
E.K. Sklyanin,
\newblock{\em The Goryachev-Chaplygin top and the method of the inverse scattering problem},
Differential geometry, Lie groups and mechanics, VI.\newblock{\em Zap. Nauchn. Sem. LOMI.}, v.133, p.236, 1984.

\bibitem{skl85}
E.K. Sklyanin,
\newblock{\em The quantum Toda chain.},
\newblock{Lecture Notes in Phys.}, v.226, p.196-293, 1985.

\bibitem{skl88}
 E.K. Sklyanin, {\em Boundary conditions for integrable
 quantum systems},
 {J. Phys. A}, v.21, p.2375, 1988.

\bibitem{skl92}
E.K. Sklyanin,
\newblock{\em Quantum Inverse Scattering Method. Selected Topics},
in {``Quantum Group and Quantum Integrable Systems''}
(Nankai Lectures in Mathematical Physics), ed.\ Mo-Lin Ge
(World Scientific, p.3--97, 1992.







\bibitem{smirn98}
F.A. Smirnov,
\newblock{\em Structure of matrix elements in quantum Toda chain},
J. Phys. A: Math. Gen. v.31, p.8953-8971, 1998.

\bibitem{ts06a}
 A. V. Tsiganov, Yu.A. Grigoryev,
\newblock{\em On the Darboux-Nijenhuis variables for the open Toda lattice},
 submittted to Vadim Kuznetsov Memorial Issue of SIGMA.


\bibitem{ts06c}
A.V. Tsiganov,
\newblock{\em Compatible Lie-Poisson brackets on Lie algebras e(3) and so(4)},
accepted to Theor. Math. Phys., 2006.

\bibitem{ts06b}
A.V. Tsiganov,
\newblock{\em On the Darboux-Nijenhuis coordinates for the
generalized open Toda lattices}, submitted to Theor. Math. Phys.,
2006.


\end{thebibliography}
\end{document}